\documentclass[sigconf]{acmart}
\setcopyright{acmlicensed}
\copyrightyear{2025}
\acmYear{2025}
\acmConference[XAIxArts 2025]{Explainable AI for the Arts Workshop 2025}{June 23, 2025}{Online}

\usepackage{graphicx, caption}
\usepackage[inkscapelatex=false]{svg}
\usepackage[export]{adjustbox}
\usepackage{booktabs}
\usepackage{url}
\usepackage{natbib}
\usepackage[font=small]{caption}
\usepackage{dblfloatfix}
\usepackage{pdfpages}
\usepackage{hyperref}

\title{{\fontfamily{lmss}\selectfont fCrit}: A Visual Explanation System for Furniture Design Creative Support}

\author{Vuong Nguyen and Gabriel Vigliensoni}
    \email{vuong.nguyen@mail.concordia.ca}
    \email{gabriel.vigliensoni@concordia.ca}
\affiliation{
  \institution{Concordia University}
  \streetaddress{1455 De Maisonneuve Blvd. West}
  \city{Montréal}
  \country{Canada}
  \postcode{H3G 1M8}}

\begin{document}

\begin{abstract}

We introduce {\fontfamily{lmss}\selectfont fCrit}, a dialogue-based AI system designed to critique furniture design with a focus on explainability. Grounded in reflective learning and formal analysis, {\fontfamily{lmss}\selectfont fCrit} employs a multi-agent architecture informed by a structured design knowledge base. We argue that explainability in the arts should not only make AI reasoning transparent but also adapt to the ways users think and talk about their designs. We demonstrate how {\fontfamily{lmss}\selectfont fCrit} scaffolds this process by adapting explanations to users' design language and awareness levels through a functional prototype. This work contributes to Human-Centred Explainable AI (\textsc{HCXAI}) in creative practice, advancing domain-specific methods for situated, dialogic, and visually grounded AI support.

\medskip

\noindent Additional Key Words and Phrases: human-centred explainable AI, mixed-initiative interaction, chatbot, domain-specific LLM, knowledge base, design critique, reflective learning, furniture design

\end{abstract}

\maketitle

\section{Introduction}

Explainability is intrinsic to design critique \cite[]{uluoglu_design_2000}.
Furniture designers commonly employ \textit{Formal Critique}---an exercise focussing on formal elements to develop formal awareness \cite[]{hannah_elements_2002}. This involves articulating tacit formal elements and analysing their relationships with design intent, often identifying both successful formal relationships and contradictions as opportunities for refinement. Based on this understanding, effective explainability must not only make formal concepts and patterns transparent, but also scaffold the designer's ability to identify and articulate these elements themselves. As recent work in Human-Centred Explainable AI (\textsc{HCXAI}) suggests \cite[]{bryan-kinns_reflections_2024, ridley_human-centered_2025}, explanations should align with the user's awareness, language, and intents.

In this paper, we present a functional prototype of {\fontfamily{lmss}\selectfont fCrit}, an AI critiquing system for furniture design that supports visual thinking through reflective and adaptive feedback. Our system is based on the premise that explainability in the arts requires interpretive alignment, dialogic responsiveness, and domain knowledge structured from established formal critique methodology \cite[]{hannah_elements_2002, wong_principles_1993}. Our contributions are twofold: (1) a theoretical framework combining reflective learning, formal critique, and a mixed-initiative creative interface that adapts mixed-initiative interaction principles for design practice; and (2) a system architecture enabling adaptive explainability grounded in visual design concepts and patterns.

\section{Related Work} 
Evaluating design solutions and enhancing designer's task performance are key functions of critiquing systems \cite[]{robbins_cognitive_2000,silverman_survey_1992}. Their architecture comprises an analyser module that matches user input against reference knowledge base, and a dialogue module for error detection \cite[]{silverman_survey_1992}. While these systems have found widespread application in design---from design sketching \cite[]{oh_intelligent_2004} to architectural layout planning \cite[]{woodbury_code_2000}, essential aspects of creativity support---feedback modality \cite[]{oh_computational_2017}, cognitive ergonomics \cite[]{tianfield_critic_2004}, and contextual awareness \cite[]{fischer_context-aware_2012}---remain neglected in their development.

Tutoring systems emerged from the need for personal support system for learners, with capabilities enhanced by large language models (LLMs) \cite[]{ait_baha_impact_2024}. Nevertheless, design critique demands more sophisticated knowledge structures than conventional error-detection approach, especially when translating designers' informal expressions and visual language.

\begin{figure}[!t]
  \includegraphics[width=1\linewidth]{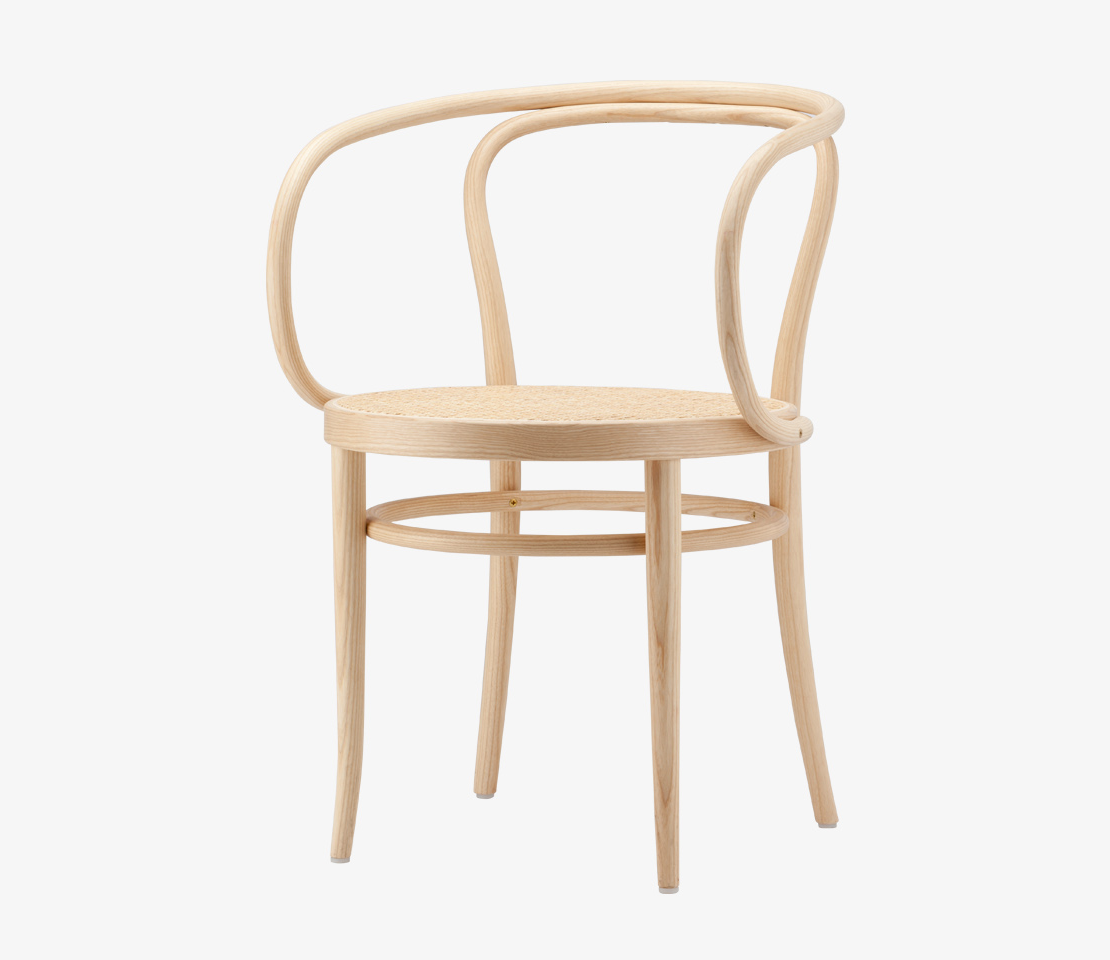}
  % \captionsetup{aboveskip=5pt}
  % \captionsetup{belowskip=-6pt}
  \captionsetup{width=1\columnwidth}
  % \captionsetup{margin={\leftmargini,33pt}}
  \caption{Gebrüder Thonet's M-209 bentwood armchair (circa 1900). The dialogue demonstration in Section 6 shows how fCrit interprets informal descriptions of its curvilinear elements (``noodle-y,'' ``playful'') and scaffolds formal design analysis.\\ 
  \text{\url{https://www.thonet.de/en/all-products/detail/209}.}}
  \Description{An image of Gebrüder Thonet's M-209 armchair, circa 1900, highlighting its elegantly curved wooden frame and flowing design. The armchair serves as a case study for our dialogue demonstration.}
  \label{fig: furnituresample}
\end{figure}

This limitation highlights the need for design-oriented learning systems that accommodate designers' natural thought processes, communication patterns, and domain-specific vocabulary. Adept design critique systems must entail competencies on \textit{adaptive design thinking} \cite[]{sharp_iterate_2019}, \textit{deep reflection} \cite[]{redstrom_making_2017}, and \textit{problem framing} \cite[]{cross_designerly_2006-1}. In the following sections, we demonstrate how {\fontfamily{lmss}\selectfont fCrit} addresses these shortcomings through a structured knowledge base and a responsive dialogic architecture, using Gebrüder Thonet’s M‑209 armchair (Fig.~\ref{fig: furnituresample}) as our illustrative case study.

\section{Theoretical Framework}
Our framework encompasses recent calls for XAI approaches that support sensemaking in creative domains \cite[]{bryan-kinns_reflections_2024, ridley_human-centered_2025}. A system must adapt its language to match the user's awareness and choice of words, and encourage deeper reflection. We integrate three frameworks that support explainability in furniture formal critique:

\subsubsection*{Reflective Learning}
Following \citet{schon_reflective_1992} and \citet{moon_handbook_2013}, critique should enhance reflection-in-action and support the articulation of tacit knowledge. Rather than prescribing solutions, {\fontfamily{lmss}\selectfont fCrit} scaffolds design reflection by mirroring insights, prompting elaboration, and constructing visual analogies.

\subsubsection*{Mixed-Initiative Creative Interface:} Inspired by \cite{deterding_mixed-initiative_2017}, rather than a unidirectional exchange, we frame and model the interaction as a co-constructive dialogue where both human and AI can take initiative towards insight discovery---the system dynamically adjusts its level of intervention and critiquing tone based on the designer's need and evolving observations. 

\subsubsection*{Formal Critique Terminology:} {\fontfamily{lmss}\selectfont fCrit} uses a domain-specific vocabulary derived from prominent formal critique literature \cite[]{hannah_elements_2002, wong_principles_1993}. The knowledge base encodes visual \textit{concepts} (e.g., line, shape, form) and \textit{patterns} (e.g., balance, contrast, unity), with multiple levels of abstraction and formal awareness. Lastly, formal critique is unconcerned with regional-stylistic or period-based labelling (e.g., Scandinavian or Victorian). This approach enables the system to interpret colloquial language and make formal connections without constraining creative freedom through stylistic designation.

\section{Critique Knowledge Base}
\begin{figure}[h!]
  \centering
  \includegraphics[width=1\columnwidth]{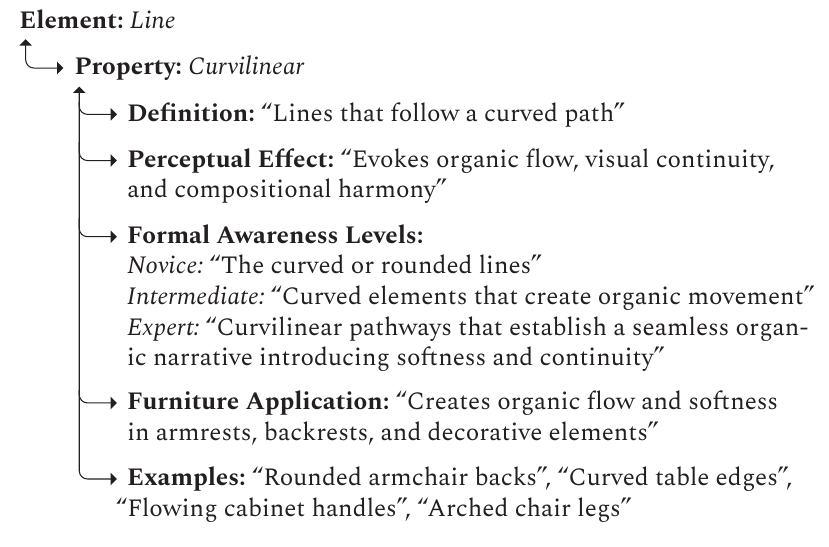}
  \caption{Hierarchical representation of Curvilinear Line in {\fontfamily{lmss}\selectfont fCrit}'s knowledge base, showing how visual concepts are structured with formal definitions, perceptual effects, awareness-tailored language, applications, and examples to support adaptive design critique.}
  \Description{A visual representation of the concept `Curvilinear Line' as organised within fCrit's knowledge base. The hierarchy includes formal definitions, perceptual effects, tailored language for different awareness levels, domain-specific furniture design applications, and examples such as rounded armchair backs and curved table edges.}
  \label{fig: embedding}
\end{figure}

Inspired by the key principle of \textit{tailored explanation} \cite{bryan-kinns_reflections_2024, ridley_human-centered_2025} and structured after formal critique methodology \cite[]{hannah_elements_2002, wong_principles_1993}, a knowledge base for formal critique is built to capture visual concepts and patterns as a structured embedding containing multiple ontological perspectives. Each embedding includes formal definitions, perceptual effects, applications, examples, and awareness-specific terminology. This structure supports adaptive explanations while ensuring design principle consistency. For instance, 
Figure~\ref{fig: embedding} shows a hierarchical representation of properties for \textit{Curvilinear Line}, which could be used to describe some aspects of the chair in Figure~\ref{fig: furnituresample}. This case example is demonstrated in Section 6 to illustrate how the system connects user descriptions like ``noodle-y'' and ``playful'' to formal concepts such as \textit{curvilinearity} and \textit{rhythmic repetition}---exemplifying how the knowledge base promotes tacit insight interpretation and visual literacy.

\section{System Architecture}

{\fontfamily{lmss}\selectfont fCrit} is implemented on n8n,\footnote{\url{https://github.com/n8n-io/n8n}} a workflow automation platform connecting modular API endpoints. Within n8n, {\fontfamily{lmss}\selectfont fCrit} orchestrates an agentic workflow that coordinates the following five specialised AI agents. 

\subsubsection*{Command Hub} Serves as the central entry point and orchestration layer; determines which agents should be activated to ensure efficient context-aware resource allocation

\subsubsection*{Design Concept Mapper} Translates user terminology into formal design critique knowledge; extracts visual \textit{concepts} (e.g., diagonal line, negative space, etc.) from the knowledge base, assigns a confidence score to each extraction, and tailors outputs according to user awareness level

\subsubsection*{Pattern Recognition Engine} Extracts visual \textit{patterns} (e.g., unity, contrast, repetition, etc.) based on user observation from the knowledge base; assigns a confidence score to each extraction

\subsubsection*{Etiquette Classifier} Determines the appropriate language mode (i.e., casual, detailed, or expert); recommends response length and formality; fosters positive emotional states by reinforcing a shared language space

\subsubsection*{Dialogue Agent} Synthesises insights from previous components to craft user-centred responses and scaffolds reflective learning progressively through three techniques: rephrasing user input using design-related language to nudge further elaboration (e.g., ``I can see why...''), asking generative questions to promote convergent thinking (e.g., ``What makes you think...?''), and creating visual analogies to stimulate divergent thinking (e.g., ``The flowing lines resemble...'').

\begin{figure}[b!]
    \centering
    \includegraphics[width=1\columnwidth]{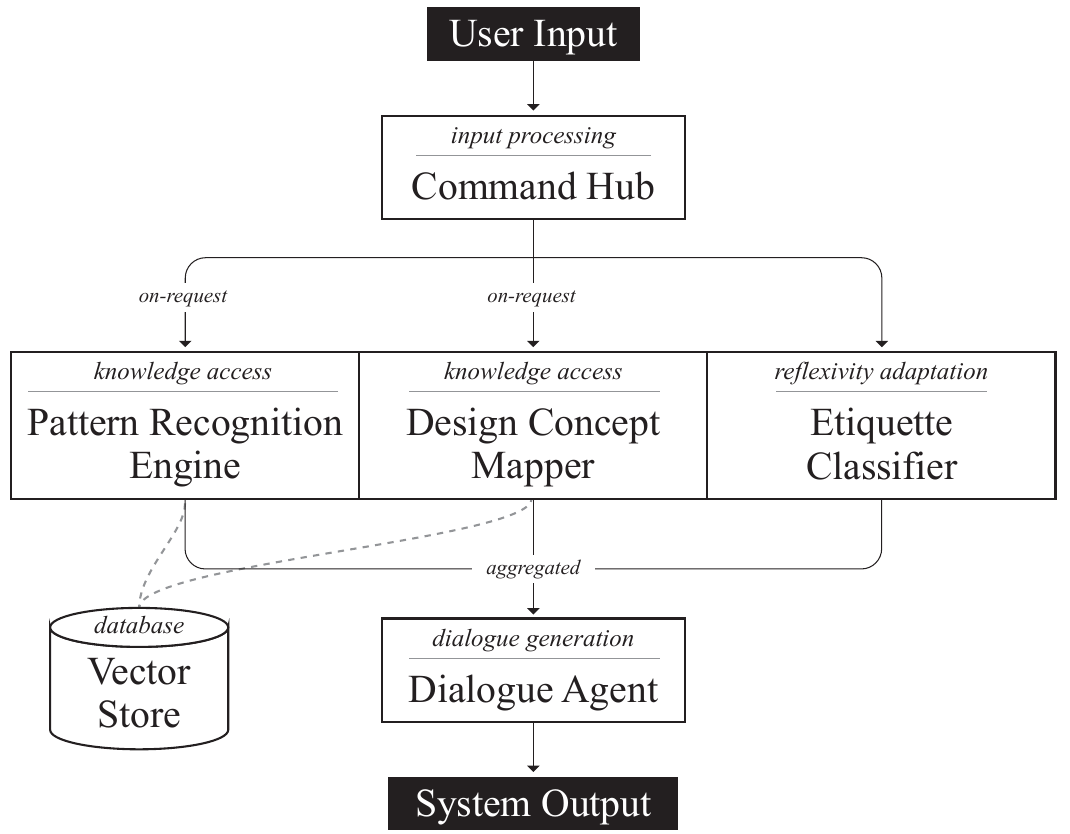}
    % \captionsetup{aboveskip=7pt}
    % \captionsetup{belowskip=-10pt}
    \caption{{\fontfamily{lmss}\selectfont fCrit} architecture: Five specialised AI agents work in a three-tiered workflow to process user input, access design knowledge, and generate furniture critiquing dialogues with adaptive levels of reflexivity. Two of them gain on-request access to a domain-specific knowledge base.}
    \label{fig: architecture}
    \Description{fCrit's architecture employs five specialised AI agents in a three-tiered workflow---input processing, on-request knowledge access to a domain-specific vector store, and dialogue generation.}
\end{figure}

As illustrated in Fig.~\ref{fig: architecture}, these agents operate in a three-tier workflow: input processing, knowledge access/adaptation, and dialogue generation. Each tier is powered by purpose-specific Claude models: Claude 3.5 Haiku for fast input processing, Claude 3 Haiku for basic knowledge retrieval, and Claude 3.7 Sonnet for the Dialogue Agent, which requires broader context and more nuanced generation.
The Design Concept Mapper and Pattern Recognition Engine agents retrieve information from the vector-based knowledge store only when relevant. This on-demand querying avoids cognitive overload, keeps feedback focused, and facilitates introspection through confidence scoring. These scores enable ongoing evaluation of system performance and knowledge base coverage.

This architecture emphasises co-creative engagement by combining modular control, distributed reasoning, and consistent prompting. It constrains the scope of each interaction to maintain clarity and progressively build shared understanding. Rather than presenting multiple perspectives simultaneously, {\fontfamily{lmss}\selectfont fCrit} guides users through one exploration at a time, thereby enhancing reflection and reducing noise.

By integrating a structured critiquing knowledge base with a retrieval-based architecture, {\fontfamily{lmss}\selectfont fCrit} enables just-in-time, context-sensitive support. Each dialogue turn adds nuance and continuity, facilitating collaboration without compromising creative agency. The system follows the designer’s lead, offering formal insights only when they enrich the current trajectory of thought.

\section{Dialogue Demonstration}
The following dialogue presents a four-turn interaction between a user and {\fontfamily{lmss}\selectfont fCrit}, using Fig.~\ref{fig: furnituresample} as the case example.

\begin{quote}

\begin{small}
\textbf{Designer:}
Hi, I like this design \textit{(with Fig.~\ref{fig: furnituresample} attached)}.

\textbf{{\fontfamily{lmss}\selectfont fCrit}:}
What specifically attracts you to this chair design? I notice it has those beautifully bent wood elements that create a continuous flowing line from the backrest to the armrests. The natural light wood tone gives it a timeless quality too.

\textbf{Designer:}
Well I'm drawn to how noodle-y it looks. It's playful yet elegant!

\textbf{{\fontfamily{lmss}\selectfont fCrit}:}
I love your description of it as ``noodle-y''! That's a great way to capture those continuous curved lines. The way the wood bends and flows without interruption creates that playful character you've noticed, while still maintaining an elegant presence. Do you think it's the smoothness of the curves or the rhythmic repetition of the lines that gives it that playful quality?

\textbf{Designer:}
I think it's in the way the curved lines traverse across different parts of the chair, tying it together to create a story. What do you think about this linear narrative?

\textbf{{\fontfamily{lmss}\selectfont fCrit}:}
That's very perceptive! Those curved lines do more than just connect physical elements---they create a visual journey through the chair's form. The way each curve flows into the next creates a sense of continuity that tells a story of movement and rhythm. It's almost as if the designer wanted us to experience the chair as a single gesture rather than separate parts. The curves also guide the eye around the entire piece, revealing how intentionally unified the design is. Do you feel this narrative quality makes the chair more inviting or more sculptural in nature?

\textbf{Designer:}
This narrative definitely invites closer inspection. There's something else about the nature of curve that feels amiable. I can't quite find the right word.

\textbf{{\fontfamily{lmss}\selectfont fCrit}:}
There's a certain warmth and hospitality in organic curves that straight lines often lack. Perhaps what you're sensing might be described as ``haptic invitation''---the way these gentle curves seem to welcome touch and interaction. The curves have a natural ergonomic quality that resonates with our bodies and movement patterns, creating an immediate sense of comfort even before physical contact.

The gradual, predictable nature of these curves also creates visual ease---our eyes can follow them without abrupt interruptions, which might contribute to that feeling of amiability you're describing. Does that resonance with human movement patterns and visual comfort capture some of what you're feeling about these curves?

\end{small}
    
\end{quote}

This interaction highlights several critiquing techniques that {\fontfamily{lmss}\selectfont fCrit} employs to interpret and translate affective and metaphorical language into formal design terms, and guide reflective inquiry while building trust and fostering positive engagement \cite[]{moon_handbook_2013}.

The opening exchange demonstrates {\fontfamily{lmss}\selectfont fCrit}'s \textit{question-first protocol}. When designer interest is ambiguous, the system prioritises understanding the designer's perspective before offering formal analysis. This is supported by \textit{response length control}, which prevents the system from overwhelming initial exchanges while still building trust and advancing critique. 

The second turn exhibits \textit{designer-centred language adaptation}, where {\fontfamily{lmss}\selectfont fCrit} focusses on the designer's natural expression (``noodle-y'') rather than immediately imposing formal terminology. The subsequent turns show \textit{grounded interpretation} where the system builds on the designer's metaphor (``linear narrative'') by introducing formal concepts (``haptic invitation'') to expand analytical scope. 

Throughout, {\fontfamily{lmss}\selectfont fCrit} uses \textit{generative questioning} to encourage deeper reflection with questions like ``Do you think it's the smoothness of the curves...?'' or ``Does that capture some of what you're feeling...?'' inviting the designer to reframe and refine their own observations. 

This dialogue demonstration exemplifies {\fontfamily{lmss}\selectfont fCrit}'s core principle of reflective learning through adaptive formal critique. Rather than imposing stylistic classifications, {\fontfamily{lmss}\selectfont fCrit}'s critique offers pragmatic formal insights that furniture designer often need while encouraging designers to develop their own analytical framework from their intuitive vocabulary.

\section{Conclusion \& Future Work}

Our immediate goal is to evaluate {\fontfamily{lmss}\selectfont fCrit} through a user study with both novice and expert designers, assessing its helpfulness, alignment with design intent, and support for reflection. Results will guide further development, particularly for deployment in educational and industry contexts, and inform adaptations for other design disciplines. This next development phase extends our core research aim: advancing AI tools that augment, rather than replace, creative judgment. By integrating visual explainability with reflective dialogue and formal analysis, {\fontfamily{lmss}\selectfont fCrit} models an XAI approach that foregrounds subjective interpretation, social interaction, and critical reflection—core dimensions of design critique.

Beyond furniture design, the system’s architecture and methodology point to broader applications across creative domains. As AI tools for designers continue to concentrate on content generation over other aspects of design practice \cite[]{shi_understanding_2023}, tools like {\fontfamily{lmss}\selectfont fCrit} may play a pivotal role in promoting design literacy, fostering problem-finding, and reshaping designer-AI collaboration. Such systems could enhance conceptual understanding, support deeper engagement with design thinking, and expand the scope of creative sensemaking in both academic and professional settings.

\bibliographystyle{ACM-Reference-Format}
\bibliography{xaixarts_positionpaper_references_2025}

\end{document}